\documentclass[pra,twocolumn,superscriptaddress,showpacs]{revtex4-1}
\usepackage{amsmath,amssymb,bm,graphicx,hyperref}

\renewcommand\d{\partial}
\newcommand\grad{\bm{\nabla}}
\newcommand\+{\dagger}
\newcommand\x{{\bm{x}}}
\newcommand\y{{\bm{y}}}
\renewcommand\k{{\bm{k}}}
\newcommand\p{{\bm{p}}}
\newcommand\q{{\bm{q}}}
\renewcommand\r{{\bm{r}}}
\newcommand\R{{\bm{R}}}
\newcommand\eps{\varepsilon}
\renewcommand\L{\mathcal{L}}

\begin{document}

\title{Super Efimov effect for mass-imbalanced systems}

\author{Sergej Moroz}
\affiliation{Department of Physics, University of Washington,
Seattle, Washington 98195, USA}
\affiliation{Department of Physics, University of Colorado,
Boulder, Colorado 80309, USA}
\affiliation{Center for Theory of Quantum Matter, University of Colorado,
Boulder, Colorado 80309, USA}
\author{Yusuke Nishida}
\affiliation{Department of Physics, Tokyo Institute of Technology,
Ookayama, Meguro, Tokyo 152-8551, Japan}

\date{July 2014}

\begin{abstract}
 We study two species of particles in two dimensions interacting by
 isotropic short-range potentials with the interspecies potential
 fine-tuned to a $p$-wave resonance.  Their universal low-energy physics
 can be extracted by analyzing a properly constructed low-energy
 effective field theory with the renormalization group method.
 Consequently, a three-body system consisting of two particles of one
 species and one of the other is shown to exhibit the super Efimov
 effect, the emergence of an infinite tower of three-body bound states
 with orbital angular momentum $\ell=\pm1$ whose binding energies obey a
 doubly exponential scaling, when the two particles are heavier than the
 other by a mass ratio greater than 4.03404 for identical bosons and
 2.41421 for identical fermions.  With increasing the mass ratio, the
 super Efimov spectrum becomes denser which would make its experimental
 observation easier.  We also point out that the Born-Oppenheimer
 approximation is incapable of reproducing the super Efimov effect, the
 universal low-energy asymptotic scaling of the spectrum.
\end{abstract}


\maketitle

\section{Introduction}
When quantum particles interact by a short-range potential with a
scattering length much larger than the potential range, they may form
universal bound states whose properties are independent of microscopic
physics~\cite{Nielsen:2001,Jensen:2004,Braaten:2006}.  Besides universal
$N$-boson bound states in one dimension~\cite{McGuire:1964} and in two
dimensions~\cite{Hammer:2004}, the most remarkable example is the Efimov
effect in three dimensions, which predicts the emergence of an infinite
tower of three-boson bound states with orbital angular momentum $\ell=0$
whose binding energies obey the universal exponential
scaling~\cite{Efimov:1970}.

\begin{table}[b]
 \caption{Comparison of the Efimov effect versus the super Efimov
 effect~\cite{Nishida:2013}.  \label{tab:comparison}}
 \begin{ruledtabular}
  \begin{tabular}{ccccc}
   & Efimov effect && Super Efimov effect & \\[2pt]\hline
   & Three bosons && Three fermions & \\
   & Three dimensions && Two dimensions & \\
   & $s$-wave resonance && $p$-wave resonance & \\
   & $\ell=0$ && $\ell=\pm1$ & \\
   & Exponential scaling && Doubly exponential scaling &
  \end{tabular}
 \end{ruledtabular}
\end{table}

Recently, new few-body universality was discovered at a $p$-wave
resonance in two dimensions~\cite{Nishida:2013}, which predicts the
emergence of an infinite tower of three-fermion bound states with
orbital angular momentum $\ell=\pm1$ whose binding energies obey the
universal doubly exponential scaling
\begin{align}
 E_n \propto \exp\bigl(-2e^{3\pi n/4+\theta}\bigr)
\end{align}
for sufficiently large $n\in\mathbb{Z}$.  It is, to the best of our
knowledge, the unique physics phenomenon exhibiting doubly exponential
scaling similarly to hyperinflation in economics~\cite{Mizuno:2002}.
This super Efimov effect summarized in Table~\ref{tab:comparison}
stimulated further theoretical studies in the hyperspherical
formalism~\cite{Volosniev:2014,Gao:2014} and its mathematical proof was
provided in Ref.~\cite{Gridnev:2014}.  On the other hand, from the
experimental perspective, the doubly exponential scaling of the binding
energies makes the experimental observation of the super Efimov spectrum
challenging.

In this article, we extend the super Efimov effect to mass-imbalanced
systems, motivated by the fact that the usual Efimov spectrum becomes
denser with increasing the mass ratio~\cite{Amado:1972,Efimov:1972}.
This advantage recently made it possible to observe up to three Efimov
resonances in ultracold atom experiments with a highly mass-imbalanced
mixture of $^6$Li and $^{133}$Cs~\cite{Pires:2014,Tung:2014}.
Correspondingly, we consider two species of particles in two dimensions
interacting by isotropic short-range potentials with the interspecies
potential fine-tuned to a $p$-wave resonance.

We first construct an effective field theory in Sec.~\ref{sec:EFT} that
properly captures the universal low-energy physics of the system under
consideration.  This low-energy effective field theory is then employed
in Sec.~\ref{sec:RG} to analyze a three-body problem consisting of two
particles of one species and one of the other with the renormalization
group method.  Consequently, such a three-body system is shown to
exhibit the super Efimov effect when the two particles are heavier than
the other by a mass ratio greater than 4.03404 for identical bosons and
2.41421 for identical fermions.  We also find that the super Efimov
spectrum indeed becomes denser with increasing the mass ratio, which
would make its experimental observation easier.  Finally, we point out
in Sec.~\ref{sec:BO} that the Born-Oppenheimer approximation is
incapable of reproducing the super Efimov effect, the universal
low-energy asymptotic scaling of the spectrum, and
Sec.~\ref{sec:summary} is devoted to the summary and conclusion of this
article.  For readers unfamiliar with our renormalization group analysis
of the low-energy effective field theory, an explicit model analysis is
also presented in the Appendix to confirm the predicted super Efimov
effect.

\section{Low-energy effective field theory}\label{sec:EFT}
Two species of particles in two dimensions interacting by isotropic
short-range potentials are described by
\begin{align}\label{eq:hamiltonian}
 & H = -\sum_{i=1,2}\int\!d\x\,\psi_i^\+(\x)\frac{\hbar^2\grad^2}{2m_i}\psi_i(\x) \notag\\
 & + \frac12\sum_{i,j=1,2}\int\!d\x d\y\,V_{ij}(|\x-\y|)
 \psi_i^\+(\x)\psi_j^\+(\y)\psi_j(\y)\psi_i(\x).
\end{align}
We assume that the interspecies potential $V_{12}(r)$ is fine-tuned to a
$p$-wave resonance while the intraspecies potentials $V_{11}(r)$ and
$V_{22}(r)$ are not.  Below we set $\hbar=1$ and denote total and
reduced masses of the two species by $M\equiv m_1+m_2$ and $\mu\equiv
m_1m_2/(m_1+m_2)$, respectively.

\begin{figure}[t]
 \includegraphics[width=0.9\columnwidth,clip]{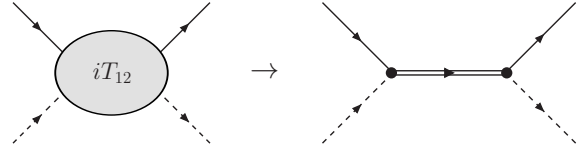}
 \caption{Low-energy limit of the resonant $p$-wave scattering in two
 dimensions reduces to the propagation of a dimer (double line) with
 energy-dependent couplings (dots) [see Eq.~(\ref{eq:resonance})].  The
 solid and dashed lines represent propagators of particles of species 1
 and 2, respectively.  \label{fig:resonance}}
\end{figure}

In order to construct an effective field theory that properly captures
the universal low-energy physics of the system described by the
Hamiltonian (\ref{eq:hamiltonian}), low-energy properties of $p$-wave
scattering in two dimensions need to be understood.
Potential-independent insights can be obtained from the effective-range
expansion for the scattering $T$-matrix in a $p$-wave
channel~\cite{Hammer:2009,tail}:
\begin{align}\label{eq:t-matrix}
 iT_{12} = \frac{2i}{\mu}\frac{2\p\cdot\q}{-\frac1{a_p}
 - \frac{4\mu\eps}{\pi}\ln\!\left(\frac{\Lambda_p}{\sqrt{-2\mu\eps}}\right)
 - \sum_{n=2}^\infty C_n(-2\mu\eps)^n}.
\end{align}
Here $\eps\equiv E-\k^2/(2M)+i0^+$ is the collision energy with $\k$
being a center-of-mass momentum, $\p$ and $\q$ are initial and final
relative momenta, respectively, while $a_p$ is the scattering area,
$\Lambda_p$ is the effective momentum, and $C_n$ are higher-order shape
parameters.  In the low-energy limit $\eps\to0$, the scattering
$T$-matrix (\ref{eq:t-matrix}) right at a $p$-wave resonance
$a_p\to\infty$ reduces to the inspiring form of
\begin{align}\label{eq:resonance}
 iT_{12} \to 2\p\cdot\q\frac{-\pi}
 {2\mu^2\ln\!\left(\frac{\Lambda_p}{\sqrt{-2\mu\eps}}\right)}
 \frac{i}{E-\frac{\k^2}{2M}+i0^+}.
\end{align}
We thus find that the last factor $iD(k)=i/[E-\k^2/(2M)+i0^+]$ has
exactly the same form as a propagator of free particle whose mass is
$M$, which indicates that the low-energy limit of the resonant $p$-wave
scattering in two dimensions is always described by the propagation of a
dimer as depicted in Fig.~\ref{fig:resonance}~\cite{resonance}.
Correspondingly, the middle factor
$(ig)^2=-\pi/[2\mu^2\ln(\Lambda_p/\sqrt{-2\mu\eps})]$ is interpreted as
a $p$-wave coupling of two scattering particles with the dimer, which
has logarithmic energy dependence and becomes small toward the
low-energy limit $\eps\to0$.

It is then straightforward to write down an effective field theory based
on the above low-energy properties of the resonant $p$-wave scattering
in two dimensions, which reads
\begin{align}\label{eq:lagrangian}
 \L_0 &= \sum_{i=1,2}\psi_i^\+\left(i\d_t+\frac{\grad^2}{2m_i}\right)\psi_i
 + \sum_{i,j=1,2}\frac{v_{ij}}2\psi_i^\+\psi_j^\+\psi_j\psi_i \notag\\
 &\quad + \sum_{\sigma=\pm}\phi_\sigma^\+\left(i\d_t+\frac{\grad^2}{2M}-\eps_0\right)\phi_\sigma \notag\\
 &\quad + g\sum_{\sigma=\pm}\phi_\sigma^\+\psi_2\left(-i\frac{m_2}{M}\overrightarrow\nabla_\sigma
 +i\frac{m_1}{M}\overleftarrow\nabla_\sigma\right)\psi_1 \notag\\
 &\quad + g\sum_{\sigma=\pm}\psi_1^\+\left(-i\frac{m_1}{M}\overrightarrow\nabla_{-\sigma}
 +i\frac{m_2}{M}\overleftarrow\nabla_{-\sigma}\right)\psi_2^\+\phi_\sigma,
\end{align}
with $\nabla_\pm\equiv\nabla_x\pm i\nabla_y$.  The couplings $v_{ij}$
represent $s$-wave components of the interspecies and intraspecies
interactions, which generally exist without fine-tunings and contribute
to low-energy scatterings.  We note that the intraspecies $s$-wave
coupling $v_{11}$ ($v_{22}$) disappears if the particle $\psi_1$
($\psi_2$) obeys the Fermi statistics.  The last three terms in the
Lagrangian density (\ref{eq:lagrangian}) represent the $p$-wave
component of the interspecies interaction, which is described by the
propagation of the dimer $\phi_\sigma$ with intrinsic angular momentum
of $\sigma=\pm1$ as observed above~\cite{resonance}.  The interspecies
$p$-wave resonance $a_p\to\infty$ is achieved by fine-tuning the bare
detuning parameter $\eps_0$ according to the relationship
$1/a_p=\Lambda^2/\pi-2\eps_0/(\mu g^2)$ with $\Lambda$ being a momentum
cutoff.

The low-energy effective field theory is not yet complete because there
are marginal three-body and four-body couplings that can be added to the
Lagrangian density
(\ref{eq:lagrangian})~\cite{Nishida:2008,Nishida:2013}.  Three-body and
four-body scatterings in our low-energy effective description are
represented by $s$-wave couplings between the particle $\psi_i$ and the
dimer $\phi_\sigma$ and between two dimers, respectively, which are
provided by
\begin{align}\label{eq:lagrangian'}
 \L' &= u_1\sum_{\sigma=\pm}\psi_1^\+\phi_\sigma^\+\phi_\sigma\psi_1
 + u_2\sum_{\sigma=\pm}\psi_2^\+\phi_\sigma^\+\phi_\sigma\psi_2 \notag\\
 &\quad + w\sum_{\sigma=\pm}\phi_\sigma^\+\phi_{-\sigma}^\+\phi_{-\sigma}\phi_\sigma
 + w'\sum_{\sigma=\pm}\phi_\sigma^\+\phi_\sigma^\+\phi_\sigma\phi_\sigma.
\end{align}
The three-body couplings $u_i$ correspond to the three-body scatterings
with total angular momentum $\ell=\pm1$, while the four-body couplings
$w$ and $w'$ correspond to the four-body scatterings with $\ell=0$ and
$\ell=\pm2$, respectively.  We note that $w'$ disappears if the $p$-wave
dimer $\phi_\sigma$ obeys the Fermi statistics.  The sum of the above
two Lagrangian densities $\L=\L_0+\L'$ now completes the low-energy
effective field theory including all marginal couplings
($v_{ij},g,u_i,w,w'$) consistent with rotation and parity symmetries and
the interspecies $p$-wave resonance, which can be employed to extract
the universal low-energy physics of the system under consideration
(\ref{eq:hamiltonian}).

\section{Renormalization group analysis}\label{sec:RG}

\subsection{Two-body sector}
\begin{figure}[t]
 \includegraphics[width=0.6\columnwidth,clip]{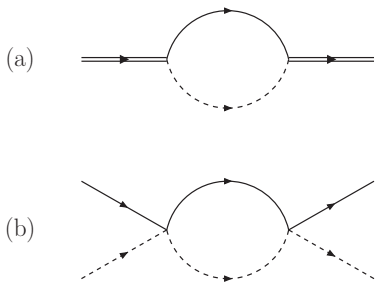}
 \caption{Feynman diagrams to renormalize the interspecies two-body
 couplings (a) $g$ and (b) $v_{12}$.  The intraspecies two-body
 couplings $v_{11}$ and $v_{22}$ are renormalized by Feynman diagrams
 similar to the one (b).  \label{fig:2-body}}
\end{figure}

The effective-range expansion for the scattering $T$-matrix indicated
that the interspecies $p$-wave coupling $g$ has logarithmic energy
dependence.  This running of the coupling is achieved in the low-energy
effective field theory (\ref{eq:lagrangian}) by its
renormalization~\cite{Nishida:2008,Nishida:2013}.  The Feynman diagram
that renormalizes $g$ is depicted in Fig.~\ref{fig:2-body}(a) and the
running of $g$ at the momentum scale $\kappa\equiv e^{-s}\Lambda$ is
governed by the renormalization group equation:
\begin{align}\label{eq:p-wave}
 \frac{dg}{ds} = -\frac{\mu^2}{\pi}g^3 \quad\Rightarrow\quad
 g^2(s) = \frac1{\frac1{g^2(0)}+\frac{2\mu^2}{\pi}s}.
\end{align}
We thus find that the interspecies $p$-wave coupling in the low-energy
limit $s=\ln\Lambda/\kappa\to\infty$ indeed becomes small
logarithmically as $g^2\to\pi/(2\mu^2s)$ in agreement with the
observation from the effective-range expansion (\ref{eq:resonance}).

Similarly, the interspecies and intraspecies $s$-wave couplings $v_{ij}$
are renormalized by a type of Feynman diagrams depicted in
Fig.~\ref{fig:2-body}(b).  The renormalization group equations that
govern the running of $v_{ij}$ and their solutions are provided by
\begin{align}
 \frac{dv_{12}}{ds} = \frac{\mu}{\pi}v_{12}^2 \quad\Rightarrow\quad
 v_{12}(s) = \frac1{\frac1{v_{12}(0)}-\frac{\mu}{\pi}s}
\end{align}
for the interspecies coupling and
\begin{align}\label{eq:s-wave}
 \frac{dv_{11}}{ds} &= \frac{m_1}{2\pi}v_{11}^2 \quad\Rightarrow\quad
 v_{11}(s) = \frac1{\frac1{v_{11}(0)}-\frac{m_1}{2\pi}s}
 \intertext{and}
 \frac{dv_{22}}{ds} &= \frac{m_2}{2\pi}v_{22}^2 \quad\Rightarrow\quad
 v_{22}(s) = \frac1{\frac1{v_{22}(0)}-\frac{m_2}{2\pi}s}
\end{align}
for the intraspecies couplings assuming the Bose statistics obeyed by
the particle field $\psi_i$.  Therefore, these $s$-wave couplings in the
low-energy limit $s\to\infty$ also become small logarithmically as
$v_{12}\to-\pi/(\mu s)$, $v_{11}\to-2\pi/(m_1s)$, and
$v_{22}\to-2\pi/(m_2s)$, all of which turn out to be negative,
indicating effective repulsion regardless of their initial signs for
$v_{ij}(0)$, i.e., attractive or repulsive potentials.

\subsection{Three-body sector}
\begin{figure*}[t]
 \includegraphics[width=1.3\columnwidth,clip]{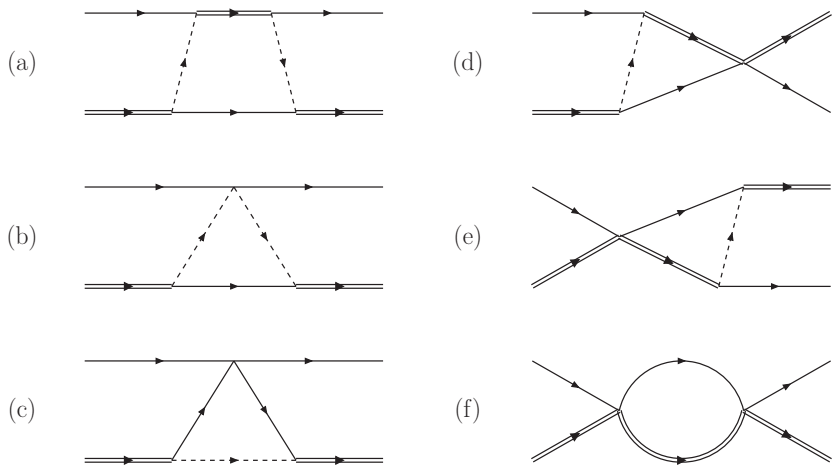}
 \caption{Feynman diagrams to renormalize the three-body coupling
 $u_1$.  \label{fig:3-body}}
\end{figure*}

We now turn to the renormalization of the three-body couplings $u_i$ in
Eq.~(\ref{eq:lagrangian'}).  Without loss of generality, we focus on the
renormalization group flow of $u_1$ because that of $u_2$ is simply
obtained by the exchange of labels $1\leftrightarrow2$.  In addition to
the contribution from the wave-function renormalization of the
$\phi_\sigma$ field, there are six distinct diagrams that renormalize
$u_1$ as depicted in Fig.~\ref{fig:3-body}.  Accordingly, after
straightforward calculations~\cite{Nishida:2008,Nishida:2013}, the
renormalization group equation that governs the running of $u_1$ is
found to be
\begin{align}\label{eq:3-body}
 \frac{du_1}{ds} &= -\frac{2\mu^2}{\pi}g^2u_1
 + \frac{8\mu^4\nu_1}{\pi m_2^2}g^4
 + \frac{2\mu^2}{\pi}g^2v_{12} \notag\\
 &\quad + \frac{4\mu^2}{\pi}g^2v_{11}\delta_{\pm+}
 \pm \frac{4\mu^2\nu_1}{\pi m_2}g^2u_1
 + \frac{\nu_1}{\pi}u_1^2,
\end{align}
where the upper (lower) sign corresponds to the case of the bosonic
(fermionic) $\psi_1$ field and $\nu_i\equiv m_iM/(m_i+M)$ is the reduced
mass of the particle of species $i$ and the dimer.  Each diagram in
Fig.~\ref{fig:3-body} contributes to the (a) second, (b) third, (c)
fourth, (d,e) fifth, and (f) sixth term in the right-hand side of
Eq.~(\ref{eq:3-body}), while its first term originates from the
wave-function renormalization of the $\phi_\sigma$ field depicted in
Fig.~\ref{fig:2-body}(a).

By substituting the low-energy asymptotic forms of the two-body
couplings $g$ and $v_{ij}$ obtained from
Eqs.~(\ref{eq:p-wave})--(\ref{eq:s-wave}), the renormalization group
equation (\ref{eq:3-body}) can be solved analytically and the three-body
coupling $u_1$ in the low-energy limit $s\to\infty$ is provided by
\begin{align}\label{eq:solution}
 su_1(s) \to \mp\frac{\pi}{m_2} - \frac{\pi\gamma}{\nu_1}\cot[\gamma(\ln s-\theta)].
\end{align}
Here $\theta$ is a nonuniversal constant depending on initial conditions
for $g$, $v_{ij}$, and $u_1$ at the microscopic scale $s\sim0$, while
$\gamma\equiv\sqrt{\nu_1^2/m_2^2-\nu_1/\mu-(4\nu_1/m_1)\delta_{\pm+}}$
is the universal exponent expressed in terms of $m_1$ and $m_2$ as
\begin{align}\label{eq:boson}
 \gamma = \frac{\sqrt{(m_1+m_2)(m_1^3-m_1^2m_2-11m_1m_2^2-5m_2^3)}}{(2m_1+m_2)m_2}
\end{align}
in the case of the bosonic $\psi_1$ field (upper sign) and
\begin{align}\label{eq:fermion}
 \gamma = \frac{(m_1+m_2)\sqrt{m_1^2-2m_1m_2-m_2^2}}{(2m_1+m_2)m_2}
\end{align}
in the case of the fermionic $\psi_1$ field (lower sign).

\begin{figure}[b]
 \includegraphics[width=0.9\columnwidth,clip]{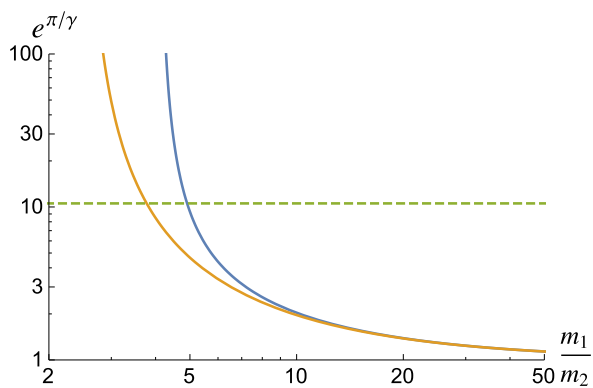}
 \caption{The universal scaling factor $e^{\pi/\gamma}$ as a function of
 the mass ratio $m_1/m_2$ for two identical bosons (upper curve) and
 fermions (lower curve) with the universal exponent $\gamma$ determined
 in Eqs.~(\ref{eq:boson}) and (\ref{eq:fermion}), respectively.  The
 horizontal dashed line indicates $e^{3\pi/4}\approx10.55$ corresponding
 to the universal scaling factor for three identical
 fermions~\cite{Nishida:2013}.  \label{fig:plot}}
\end{figure}

When $\gamma$ is real, the low-energy asymptotic solution
(\ref{eq:solution}) for $su_1$ is a periodic function of $\ln s$ and
diverges at $\ln s_n=\pi n/\gamma+\theta$.  These divergences in the
renormalization group flow of the three-body coupling $u_1$ indicate the
existence of an infinite tower of characteristic energy scales
$E_n\propto\kappa_n^2=e^{-2s_n}\Lambda^2$ in the three-body system
consisting of two particles of species 1 and another particle of species
2 with total angular momentum $\ell=\pm1$.  As was confirmed in
Ref.~\cite{Nishida:2013}, these energy scales correspond to binding
energies of the three particles, which leads to the super Efimov
spectrum
\begin{align}\label{eq:spectrum}
 E_n \propto \exp\bigl(-2e^{\pi n/\gamma+\theta}\bigr)
\end{align}
for sufficiently large $n\in\mathbb{Z}$.  This super Efimov effect
emerges when the majority species 1 is heavier than the minority species
2 and the critical mass ratio is found to be $m_1/m_2=4.03404$ from
Eq.~(\ref{eq:boson}) when the two particles are identical bosons and
$m_1/m_2=2.41421$ from Eq.~(\ref{eq:fermion}) when the two particles are
identical fermions.  In both cases, the universal exponent $\gamma$
increases monotonously with increasing the mass ratio $m_1/m_2$, which
makes the super Efimov spectrum (\ref{eq:spectrum}) denser as seen in
Fig.~\ref{fig:plot} for the logarithmic energy ratio
$\ln E_{n+1}/\ln E_n\to e^{\pi/\gamma}$ determined by the universal
scaling factor.

So far we have considered the most general case where interspecies and
intraspecies $s$-wave interactions $v_{ij}$ exist when they are
possible.  For the purpose of examining the Born-Oppenheimer
approximation in the succeeding section, it is more convenient to
consider the simplest case where all $s$-wave interactions are
artificially switched off.  By setting $v_{ij}=0$ in the renormalization
group equation (\ref{eq:3-body}), the universal exponent $\gamma$ in the
low-energy asymptotic solution (\ref{eq:solution}) for the three-body
coupling $u_1$ is modified to
\begin{align}\label{eq:exponent}
 \gamma = \frac{\nu_1}{m_2} = \frac{m_1(m_1+m_2)}{(2m_1+m_2)m_2}.
\end{align}
Because $\gamma$ is always real without $s$-wave interactions, the super
Efimov effect emerges for any mass ratio $m_1/m_2$.  In particular, the
super Efimov spectrum (\ref{eq:spectrum}) becomes independent of whether
the two particles are identical bosons or fermions.  The super Efimov
effect predicted in this simple case is also confirmed with an explicit
model analysis in the Appendix.

\begin{figure*}[t]
 \includegraphics[width=1.3\columnwidth,clip]{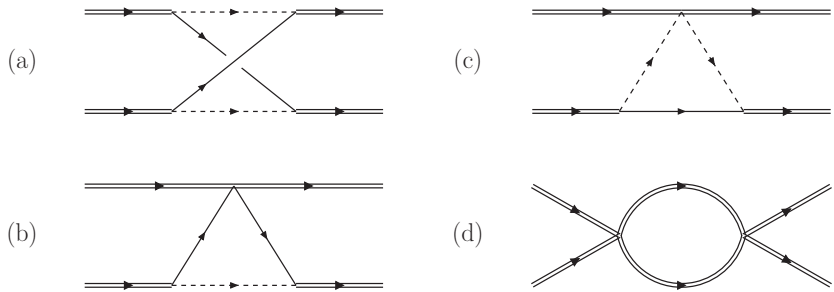}
 \caption{Feynman diagrams to renormalize the four-body couplings $w$
 and $w'$.  \label{fig:4-body}}
\end{figure*}

\subsection{Four-body sector}
We then turn to the renormalization of the four-body couplings $w$ and
$w'$ in Eq.~(\ref{eq:lagrangian'}).  In addition to the contribution
from the wave-function renormalization of the $\phi_\sigma$ field, there
are four distinct diagrams that renormalize $w$ and $w'$ as depicted in
Fig.~\ref{fig:4-body}.  Accordingly, after straightforward
calculations~\cite{Nishida:2008,Nishida:2013}, the renormalization group
equations that govern the running of $w$ and $w'$ are found to be
\begin{align}\label{eq:4-body}
 \frac{dw}{ds} &= -\frac{4\mu^2}{\pi}g^2w
 + [(\pm1)_1+(\pm1)_2]\frac{4\mu^3}{\pi}g^4 \notag\\
 &\quad + \frac{2\mu^2}{\pi}g^2u_1 + \frac{2\mu^2}{\pi}g^2u_2 + \frac{M}{\pi}w^2
\end{align}
and
\begin{align}\label{eq:4-body'}
 \frac{dw'}{ds} &= -\frac{4\mu^2}{\pi}g^2w'
 + [(\pm1)_1+(\pm1)_2]\frac{2\mu^3}{\pi}g^4 \notag\\
 &\quad + \frac{2\mu^2}{\pi}g^2u_1 + \frac{2\mu^2}{\pi}g^2u_2 + \frac{M}{\pi}w'^2.
\end{align}
Here the upper (lower) sign in $(\pm1)_i$ corresponds to the case of the
bosonic (fermionic) $\psi_i$ field and Eq.~(\ref{eq:4-body'}) assumes
the Bose statistics obeyed by the $p$-wave dimer field $\phi_\sigma$.
Each diagram in Fig.~\ref{fig:4-body} contributes to the (a) second, (b)
third, (c) fourth, and (d) fifth terms in the right-hand sides of
Eqs.~(\ref{eq:4-body}) and (\ref{eq:4-body'}), while their first terms
originate from the wave-function renormalization of the $\phi_\sigma$
field depicted in Fig.~\ref{fig:2-body}(a).

While the renormalization group flows of the four-body couplings $w$ and
$w'$ can be studied numerically~\cite{Nishida:2013}, we defer these
analyses to a future work.

\section{Born-Oppenheimer approximation}\label{sec:BO}
It is well known that the Born-Oppenheimer approximation provides
elementary and intuitive understanding of the usual Efimov
effect~\cite{Fonseca:1979}.  Therefore, it is worthwhile to examine
whether the Born-Oppenheimer approximation is useful as well to
understand the super Efimov effect.

In our system under consideration (\ref{eq:hamiltonian}), the three-body
wave function $\Psi(\R,\r)$ describing two particles of species 1
located at $\pm\R/2$ and another particle of species 2 located at $\r$
in the center-of-mass frame satisfies the Schr\"odinger equation:
\begin{align}
 & \left[-\frac{\grad_\R^2}{m_1}-\frac{\grad_\r^2}{2m}+V_{11}(R)
 +V_{12}(r_+)+V_{12}(r_-)\right]\Psi(\R,\r) \notag\\
 &= E\Psi(\R,\r),
\end{align}
where $\r_\pm\equiv\r\pm\R/2$ are interspecies separations and
$m\equiv2m_1m_2/(2m_1+m_2)$ reduces to $m_2$ at a large mass ratio
$m_1/m_2\gg1$.  The Born-Oppenheimer approximation is based on the
factorized wave function
\begin{align}\label{eq:wavefunction}
 \Psi(\R,\r) = \Phi(\R)\varphi(\R;\r),
\end{align}
where the wave function $\varphi(\R;\r)$ for the light particle
satisfies
\begin{align}\label{eq:light}
 \left[-\frac{\grad_\r^2}{2m_2}+V_{12}(r_+)+V_{12}(r_-)\right]\varphi(\R;\r)
 = \eps(R)\varphi(\R;\r),
\end{align}
with fixed locations of the two heavy particles, and the wave function
$\Phi(\R)$ for the two heavy particles in turn satisfies
\begin{align}\label{eq:heavy}
 \left[-\frac{\grad_\R^2}{m_1}+V_{11}(R)+\eps(R)\right]\Phi(\R) = E\Phi(\R),
\end{align}
with an effective potential $\eps(R)$ generated by the light particle.
Corrections to this Schr\"odinger equation (\ref{eq:heavy}) scale as
$\sim1/m_1$ and thus they are usually negligible compared to
$\eps(R)\sim1/m_2$ at a large mass ratio $m_1/m_2\gg1$.  For simplicity,
we also neglect the intraspecies potential $V_{11}(R)\to0$ and consider
only the $p$-wave component of the interspecies potential $V_{12}(r)$.

The Schr\"odinger equation (\ref{eq:light}) with the binding energy
$\eps(R)\equiv-\kappa^2/(2m_2)$ potentially admits four bound state
solutions for the light particle, whose wave functions outside the
potential range $V_{12}(r_\pm)\to0$ are expressed as
\begin{align}
 \varphi_\pm^x(\R;\r) &= K_1(\kappa r_+)\cos[\arg(\r_+)-\arg(\R)] \notag\\
 &\quad \mp K_1(\kappa r_-)\cos[\arg(\r_-)-\arg(\R)]
\end{align}
and
\begin{align}
 \varphi_\pm^y(\R;\r) &= K_1(\kappa r_+)\sin[\arg(\r_+)-\arg(\R)] \notag\\
 &\quad \mp K_1(\kappa r_-)\sin[\arg(\r_-)-\arg(\R)].
\end{align}
We note that $\varphi_+^{x,y}(\R;\r)$ [$\varphi_-^{x,y}(\R;\r)$] are
even (odd) under the exchange of the two heavy particles $\R\to-\R$.
The interspecies $p$-wave resonance is achieved by imposing the boundary
condition on the light-particle wave function
$\varphi(\R;\r)\propto1/r_\pm+O(r_\pm^3)$ at a short distance
$r_\pm\sim1/\Lambda\ll1/\kappa,R$, which leads to
\begin{align}
 \ln(\Lambda/\kappa) = \pm[K_0(\kappa R)+K_2(\kappa R)]
\end{align}
for $\varphi_\pm^x(\R;\r)$ and
\begin{align}
 \ln(\Lambda/\kappa) = \pm[K_0(\kappa R)-K_2(\kappa R)]
\end{align}
for $\varphi_\pm^y(\R;\r)$.  Because of $K_2(\kappa R)>K_0(\kappa R)>0$,
these boundary conditions can be satisfied only for $\varphi_+^x(\R;\r)$
and $\varphi_-^y(\R;\r)$ and their binding energies are found to have
the same asymptotic form of
\begin{align}\label{eq:potential}
 \eps_\pm(R) = -\frac{\kappa_\pm^2}{2m_2} \to -\frac1{m_2R^2\ln(R\Lambda)}
\end{align}
for large separation $R\Lambda\to\infty$ between the two heavy
particles.

We now solve the Schr\"odinger equation (\ref{eq:heavy}) for the two
heavy particles whose wave function can be taken as
$\Phi(\R)=e^{i\ell\arg(\R)}\Phi_\ell(R)$ with $\ell$ corresponding to
the total angular momentum of the three particles.  We first consider an
$\ell=0$ channel in which bound states are most favored due to the
absence of centrifugal barrier.  Because the total wave function
(\ref{eq:wavefunction}) has to be symmetric (antisymmetric) under the
exchange of the two heavy particles $\R\to-\R$ when they are identical
bosons (fermions), only $\varphi_+^x(\R;\r)$ [$\varphi_-^y(\R;\r)$] is
allowed for the light-particle wave function $\varphi(\R;\r)$.  Then the
Schr\"odinger equation (\ref{eq:heavy}) with the effective potential
$\eps_+(R)$ [$\eps_-(R)$] obtained in Eq.~(\ref{eq:potential}) leads to
an infinite tower of bound states whose binding energies scale
as~\cite{Gao:2014,Efremov:2014}
\begin{align}\label{eq:approximation}
 E_n^\mathrm{(BO)} \propto \exp\!\left(-\frac{m_2\pi^2}{2m_1}n^2\right)
\end{align}
for sufficiently large $n\in\mathbb{Z}$ regardless of whether the two
heavy particles are identical bosons or fermions.  On the other hand,
for higher partial-wave channels $\ell\neq0$, the low-energy asymptotic
scaling of the spectrum (\ref{eq:approximation}) is terminated around
$E\propto e^{-(2/\ell^2)m_1/m_2}$, where the centrifugal barrier
overcomes the effective potential (\ref{eq:potential}).

The resulting spectrum from the Born-Oppenheimer approximation differs
from the super Efimov spectrum (\ref{eq:spectrum}) with the universal
exponent (\ref{eq:exponent}) at a large mass ratio $m_1/m_2\gg1$,
\begin{align}\label{eq:largemass}
 E_n \propto \exp\bigl(-2e^{(2m_2/m_1)\pi n+\theta}\bigr),
\end{align}
which is the true low-energy asymptotic scaling of the spectrum as was
shown in the preceding section.  In addition, the Born-Oppenheimer
spectrum (\ref{eq:approximation}) appears in an $\ell=0$ channel, while
the super Efimov spectrum (\ref{eq:largemass}) appears in $\ell=\pm1$
channels and our analysis predicts no accumulation of infinite bound
states toward zero energy in other partial-wave channels.  Therefore, we
conclude that the Born-Oppenheimer approximation for three-body systems
with $p$-wave resonant interactions in two dimensions is incapable of
reproducing the true low-energy asymptotic scaling of the spectrum even
at a large mass ratio.  This failure of the Born-Oppenheimer
approximation may be understood in the following
way~\cite{Petrov:2014}.  When the two heavy particles are separated by a
distance $R$, their characteristic time scale is $\sim m_1R^2$, while
that of the light particle is set by the inverse of its binding energy,
$\sim m_2R^2\ln(R\Lambda)$, from Eq.~(\ref{eq:potential}).  Therefore,
even at a large mass ratio, the light particle cannot adiabatically
follow the motion of the two heavy particles for sufficiently large
separation $R\Lambda\gtrsim e^{m_1/m_2}$ where the Born-Oppenheimer
approximation fails.  This argument, however, leaves the possibility
that the resulting spectrum (\ref{eq:approximation}) may appear as an
intermediate scaling for $|E|\gtrsim e^{-2m_1/m_2}\Lambda^2/\mu$.

\section{Summary and conclusion}\label{sec:summary}
In this article, we extended the super Efimov effect to mass-imbalanced
systems (\ref{eq:hamiltonian}) where two species of particles in two
dimensions interact by isotropic short-range potentials with the
interspecies potential fine-tuned to a $p$-wave resonance.  Their
universal low-energy physics can be extracted by analyzing a properly
constructed low-energy effective field theory with the renormalization
group method~\cite{Nishida:2008,Nishida:2013}.  Consequently, a
three-body system consisting of two particles of one species and one of
the other is shown to exhibit the super Efimov spectrum
\begin{align}
 E_n \propto \exp\bigl(-2e^{\pi n/\gamma+\theta}\bigr)
\end{align}
for sufficiently large $n\in\mathbb{Z}$, when the two particles are
heavier than the other by a mass ratio greater than 4.03404 for
identical bosons [see Eq.~(\ref{eq:boson})] and 2.41421 for identical
fermions [see Eq.~(\ref{eq:fermion})].  In particular, we found that the
universal exponent $\gamma$ increases monotonously with increasing the
mass ratio which makes the super Efimov spectrum denser and thus its
experimental observation would become easier with ultracold atoms.  For
example, a highly mass-imbalanced mixture of $^6$Li and $^{133}$Cs with
their interspecies $p$-wave Feshbach resonances being
observed~\cite{Repp:2013} has the universal exponent $\gamma\approx10.7$
corresponding to the logarithmic energy ratio of
$\ln E_{n+1}/\ln E_n\to e^{\pi/\gamma}\approx1.34$, which is
significantly reduced compared to $e^{\pi/\gamma}\approx10.55$ with
$\gamma=4/3$ for three identical fermions~\cite{Nishida:2013}.

We also pointed out that the Born-Oppenheimer approximation is incapable
of reproducing the super Efimov effect, the universal low-energy
asymptotic scaling of the spectrum, even at a large mass ratio for
three-body systems with $p$-wave resonant interactions in two
dimensions.  The possible reason for this failure of the Born-Oppenheimer
approximation was elucidated, while the possibility for the resulting
spectrum (\ref{eq:approximation}) to appear as an intermediate scaling
and then crossover to the asymptotic super Efimov scaling remains to be
elucidated in a future work.

\acknowledgments
We acknowledge many useful discussions with Yvan Castin, Vitaly Efimov,
Dmitry S.~Petrov, Dam T.~Son, and participants in the INT Program on
``Universality in Few-Body Systems: Theoretical Challenges and New
Directions.''  This work was supported by U.S.\ DOE Grant
No.\ DE-FG02-97ER-41014, NSF Grant No.\ DMR-1001240, and JSPS KAKENHI
Grant No.\ 25887020.  Numerical calculations reported in the Appendix
were performed at the University of Washington Hyak cluster funded by
NSF MRI Grant No.\ PHY-0922770, the Janus supercomputer funded by NSF
Grant No.\ CNS-0821794 and the University of Colorado Boulder, and the
YITP computer facility at Kyoto University.

\appendix*
\section{Model confirmation of the super Efimov effect}
The above predictions from our renormalization group analysis of the
low-energy effective field theory are all strict as well as universal
because we do not need to specify the forms of interspecies and
intraspecies potentials in the Hamiltonian (\ref{eq:hamiltonian}).
However, since some readers may be unfamiliar with our approach, we also
present an explicit model analysis to confirm the predicted super Efimov
effect by extending that in Ref.~\cite{Nishida:2013} to mass-imbalanced
systems.

\begin{figure}[t]
 \includegraphics[width=0.9\columnwidth,clip]{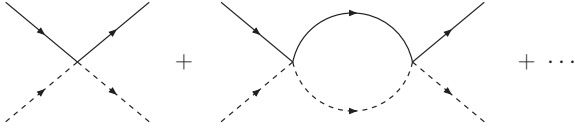}
 \caption{Feynman diagrams representing the two-body scattering
 $T$-matrix (\ref{eq:T_2-body}).  \label{fig:T_2-body}}
\end{figure}

For simplicity, we neglect the intraspecies potentials
$V_{11}(r),V_{22}(r)\to0$ and consider only the $p$-wave component of
the interspecies potential $V_{12}(r)$, which is assumed to be in a
separable form of
\begin{align}
 H &= \sum_{i=1,2}\int\!\frac{d\k}{(2\pi)^2}\,
 \frac{\k^2}{2m_i}\,\psi_i^\+(\k)\psi_i(\k) \notag\\
 &\quad - v_p\sum_{\sigma=\pm}\int\!\frac{d\k d\p d\q}{(2\pi)^6}\,
 \chi_{-\sigma}(\q)\chi_\sigma(\p)\,\psi_1^\+\!\left(\frac{m_1}{M}\k+\q\right) \notag\\
 &\qquad \times \psi_2^\+\!\left(\frac{m_2}{M}\k-\q\right)
 \psi_2\!\left(\frac{m_2}{M}\k-\p\right)\psi_1\!\left(\frac{m_1}{M}\k+\p\right),
\end{align}
with the $p$-wave form factor
$\chi_\pm(\p)\equiv(p_x\pm ip_y)e^{-\p^2/(2\Lambda^2)}$ providing the
momentum cutoff $\Lambda$.  By summing an infinite series of Feynman
diagrams depicted in Fig.~\ref{fig:T_2-body}, the scattering $T$-matrix
for this model potential is computed as
\begin{align}\label{eq:T_2-body}
 iT_{12} = \frac{2i}{\mu}\frac{2\p\cdot\q\,e^{-(\p^2+\q^2)/(2\Lambda^2)}}
 {\frac2{\mu v_p}-\frac{\Lambda^2}{\pi}-\frac{2\mu\eps}{\pi}\,
 e^{-2\mu\eps/\Lambda^2}E_1\!\left(-\frac{2\mu\eps}{\Lambda^2}\right)},
\end{align}
where $E_1(w)\equiv\int_w^\infty\!dt\,e^{-t}/t$ is the first-order
exponential integral.  The interspecies $p$-wave resonance
$a_p\to\infty$ is achieved by fine-tuning the bare $p$-wave coupling
$v_p$ according to the relationship $1/a_p=\Lambda^2/\pi-2/(\mu v_p)$,
which is obtained by comparing the computed scattering $T$-matrix
(\ref{eq:T_2-body}) on shell with the effective-range expansion
(\ref{eq:t-matrix}).

\begin{figure*}[t]
 \includegraphics[width=1.45\columnwidth,clip]{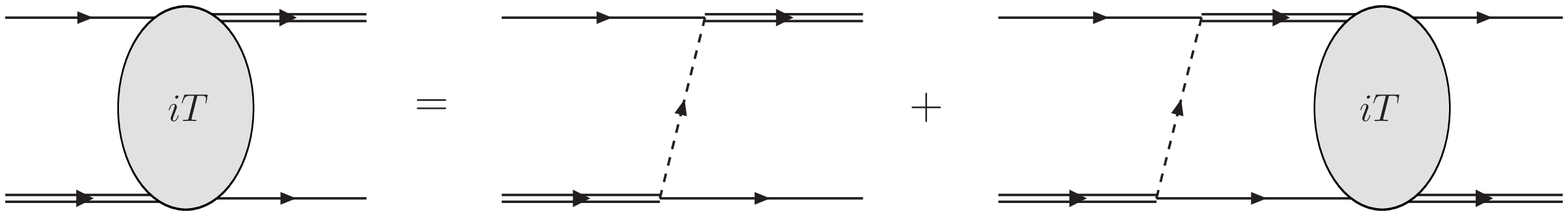}
 \caption{Feynman diagrams representing the three-body scattering
 $T$-matrix (\ref{eq:T_3-body}).  \label{fig:T_3-body}}
\end{figure*}

We are now ready to analyze a three-body problem consisting of two
particles of species 1 and another particle of species 2 right at a
$p$-wave resonance $a_p\to\infty$ in two dimensions.  Their scattering
$T$-matrix satisfies the Skorniakov--Ter-Martirosian-type integral
equation depicted in Fig.~\ref{fig:T_3-body}, which is expressed in the
center-of-mass frame as
\begin{widetext}
\begin{align}\label{eq:T_3-body}
 & T_{\sigma\sigma'}(E;\p,\p') = \pm 2\mu
 \frac{e^{-\frac{M^2+m_1^2}{2M^2}\frac{\p^2+\p'^2}{\Lambda^2}
 -\frac{2m_1}{M}\frac{\p\cdot\p'}{\Lambda^2}}}
 {\p^2+\p'^2+\frac{2m_1}{M}\p\cdot\p'-2\mu E-i0^+}
 \left(\frac{m_1}{M}\p+\p'\right)_{-\sigma}
 \left(\p+\frac{m_1}{M}\p'\right)_{\sigma'} \notag\\
 &\quad \pm \int\!\frac{d\q}{\pi}
 \frac{e^{-\frac{M^2+m_1^2}{2M^2}\frac{\p^2+\q^2}{\Lambda^2}
 -\frac{2m_1}{M}\frac{\p\cdot\q}{\Lambda^2}}}
 {\p^2+\q^2+\frac{2m_1}{M}\p\cdot\q-2\mu E-i0^+}
 \frac{\left(\frac{m_1}{M}\p+\q\right)_{-\sigma}
 \sum_{\tau=\pm}\left(\p+\frac{m_1}{M}\q\right)_\tau T_{\tau\sigma'}(E;\q,\p')}
 {\left(\frac{M^2-m_1^2}{M^2}\q^2-2\mu E-i0^+\right)
 e^{\frac{M^2-m_1^2}{M^2}\frac{\q^2}{\Lambda^2}-\frac{2\mu E+i0^+}{\Lambda^2}}
 E_1\!\left(\frac{M^2-m_1^2}{M^2}\frac{\q^2}{\Lambda^2}
 -\frac{2\mu E+i0^+}{\Lambda^2}\right)},
\end{align}
\end{widetext}
where the upper (lower) sign corresponds to the case of the bosonic
(fermionic) $\psi_1$ field and $\p$ ($\p'$) is an initial (final)
momentum of a particle of species 1 with respect to the other two
particles scattering with an orbital angular momentum of
$\sigma\ (\sigma')=\pm1$.  When the collision energy $E$ approaches the
binding energy $E\to-\kappa^2/\mu<0$, the above scattering $T$-matrix
factorizes as
$T_{\sigma\sigma'}(E;\p,\p')\to Z_\sigma(\p)Z_{\sigma'}^*(\p')/(E+\kappa^2/\mu)$
and the resulting residue function $Z_\sigma(\p)$ satisfies
\begin{align}\label{eq:residue}
 & Z_\sigma(\p) = \pm\int\!\frac{d\q}{\pi}
 \frac{e^{-\frac{M^2+m_1^2}{2M^2}\frac{\p^2+\q^2}{\Lambda^2}
 -\frac{2m_1}{M}\frac{\p\cdot\q}{\Lambda^2}}}
 {\p^2+\q^2+\frac{2m_1}{M}\p\cdot\q+2\kappa^2}\ \times \notag\\
 & \frac{\left(\frac{m_1}{M}\p+\q\right)_{-\sigma}
 \sum_{\tau=\pm}\left(\p+\frac{m_1}{M}\q\right)_\tau Z_\tau(\q)}
 {\left(\frac{M^2-m_1^2}{M^2}\q^2+2\kappa^2\right)
 e^{\frac{M^2-m_1^2}{M^2}\frac{\q^2}{\Lambda^2}+\frac{2\kappa^2}{\Lambda^2}}
 E_1\!\left(\frac{M^2-m_1^2}{M^2}\frac{\q^2}{\Lambda^2}
 +\frac{2\kappa^2}{\Lambda^2}\right)}.
\end{align}
It is easy to see that $Z_+(\p)=e^{i(\ell-1)\arg(\p)}z_+(p)$ couples to
$Z_-(\p)=e^{i(\ell+1)\arg(\p)}z_-(p)$ with $\ell$ corresponding to the
total angular momentum of the three particles.  Below we focus on an
$\ell=+1$ channel in which the super Efimov effect was shown to emerge,
while solutions in an $\ell=-1$ channel are simply obtained by the
exchange of labels $+\leftrightarrow-$.

The two coupled integral equations (\ref{eq:residue}) can be solved
analytically in the low-energy limit $\kappa/\Lambda\to0$ with the
leading-logarithm
approximation~\cite{Son:1999,Levinsen:2008,Nishida:2013}.  We assume
that the integral is dominated by the region $\kappa\ll q\ll\Lambda$ and
split the integral into two parts, $\kappa\ll q\ll p$ and
$p\ll q\ll\Lambda$, where a sum of $p$ and $q$ in the integrand is
replaced with whichever is larger.  Accordingly, Eq.~(\ref{eq:residue})
is simplified to
\begin{subequations}
\begin{align}
 \pm\frac{z_+(p)}{\gamma} &= \int_\kappa^p\!\frac{dq}{q}\,\frac{z_+(q)}{\ln\Lambda/q}
 + \int_p^{\epsilon\Lambda}\!\frac{dq}{q}\,\frac{z_+(q)+z_-(q)}{\ln\Lambda/q}, \\
 \pm\frac{z_-(p)}{\gamma} &= \int_\kappa^p\!\frac{dq}{q}\,\frac{z_+(q)}{\ln\Lambda/q},
\end{align}
\end{subequations}
where $\gamma\equiv Mm_1/(M^2-m_1^2)$ coincides with the universal
exponent (\ref{eq:exponent}) without $s$-wave interactions and
$\epsilon<1$ is a positive constant.  By changing variables to
$P\equiv\ln\ln\Lambda/p$ and $Q\equiv\ln\ln\Lambda/q$ and defining
$\lambda\equiv\ln\ln\Lambda/\kappa$, $\eta\equiv\ln\ln1/\epsilon$, and
$\zeta_\pm(P)\equiv z_\pm(p)$, we obtain
\begin{subequations}
\begin{align}
 \pm\frac{\zeta_+(P)}{\gamma} &= \int_P^\lambda\!dQ\,\zeta_+(Q)
 + \int_\eta^P\!dQ\,[\zeta_+(Q)+\zeta_-(Q)], \\
 \pm\frac{\zeta_-(P)}{\gamma} &= \int_P^\lambda\!dQ\,\zeta_+(Q).
\end{align}
\end{subequations}
These two coupled integral equations are solved by~\cite{Nishida:2013}
\begin{subequations}
\begin{align}
 \zeta_+(P) &= \cos[\mp\gamma(P-\lambda)], \\
 \zeta_-(P) &= \sin[\mp\gamma(P-\lambda)],
\end{align}
\end{subequations}
provided that the boundary condition $\zeta_+(\eta)=\zeta_-(\eta)$ is
satisfied.  This boundary condition leads to an infinite tower of
allowed binding energies $\lambda_n=\pi n/\gamma+\theta$ with
$n\in\mathbb{Z}$ for any mass ratio $m_1/m_2$ regardless of whether the
two particles are identical bosons or fermions, which indeed confirms
the predicted super Efimov effect (\ref{eq:spectrum}).

\begin{table}[b]
 \caption{Lowest seventeen three-body binding energies
 $E_n=-\kappa_n^2/\mu$ obtained from Eq.~(\ref{eq:residue}) for
 $\ell=\pm1$, $m_1/m_2=20$, and two identical bosons (upper sign).  The
 logarithmic energy ratios asymptotically approach the universal scaling
 factor $e^{\pi/\gamma}\approx1.358905074$ with $\gamma=420/41$
 determined in Eq.~(\ref{eq:exponent}).  \label{tab:binding}}
 \begin{ruledtabular}
  \begin{tabular}{ccccccc}
   & $n$ && $\ln(\Lambda/\kappa_n)$ &&
   $\ln(\Lambda/\kappa_n)/\ln(\Lambda/\kappa_{n-1})$ & \\[2pt]\hline
   & 0 && 0.84492 && --- & \\
   & 1 && 1.4017\phantom{0} && 1.6590 & \\
   & 2 && 2.5612\phantom{0} && 1.8272 & \\
   & 3 && 4.3083\phantom{0} && 1.6821 & \\
   & 4 && 6.5930\phantom{0} && 1.5303 & \\
   & 5 && 9.5792\phantom{0} && 1.4529 & \\
   & 6 && 13.513\phantom{0} && 1.4107 & \\
   & 7 && 18.740\phantom{0} && 1.3868 & \\
   & 8 && 25.742\phantom{0} && 1.3736 & \\
   & 9 && 35.177\phantom{0} && 1.3665 & \\
   & 10 && 47.939\phantom{0} && 1.3628 & \\
   & 11 && 65.240\phantom{0} && 1.3609 & \\
   & 12 && 88.720\phantom{0} && 1.3599 & \\
   & 13 && 120.61\phantom{0} && 1.3594 & \\
   & 14 && 163.92\phantom{0} && 1.3591 & \\
   & 15 && 222.77\phantom{0} && 1.3590 & \\
   & 16 && 302.73\phantom{0} && 1.3589 & \\[2pt]\hline
   & $\infty$ && --- && \ \ \ \ \ \ \ \,1.358905074 & \\
  \end{tabular}
 \end{ruledtabular}
\end{table}

We also solved the two coupled integral equations (\ref{eq:residue})
numerically with $\ell=\pm1$ at mass ratios of $m_1/m_2=5$, $10$, and
$20$ and observed that the obtained binding energies asymptotically
approach the predicted doubly exponential scaling for each mass ratio.
See Table~\ref{tab:binding} for the obtained binding energies at
$m_1/m_2=20$ for two identical bosons corresponding to the upper sign in
Eq.~(\ref{eq:residue}).


\begin{thebibliography}{99}
\bibitem{tail}
The form of the effective-range expansion up to the second order in
$k\equiv\sqrt{2\mu\eps}$ is not affected as long as the underlying
potential $V_{12}(r)$ vanishes faster than $r^{-6}$ at $r\to\infty$.
For the van der Waals potential $V_{12}(r)\to-C_6/r^6$, the denominator
of Eq.~(\ref{eq:t-matrix}) suffers a logarithmic correction of
$-[\pi C_6/(32a_p^2)]\,k^2\ln|k|$, which, however, disappears right at
the $p$-wave resonance $a_p\to\infty$ and thus does not affect the
conclusion of this article.
\end{thebibliography}
\end{document}